\documentclass{eas}
\usepackage{graphicx}
\begin{document}

\title{A SAURON study of dwarf elliptical galaxies in the Virgo Cluster} 
\author{Agnieszka Ry\'{s}} \address{Instituto de Astrof\'{i}sica de Canarias (IAC), E-38200 La Laguna, Tenerife, Spain} \secondaddress{Depto. Astrof\'{i}sica, Universidad de La Laguna (ULL), E-38206 La Laguna, Tenerife, Spain}
\author{Jes\'{u}s Falc\'{o}n-Barroso} \sameaddress{1, 2}
\author{Mina Koleva}  \sameaddress{1, 2}

%

%
\begin{abstract}
Dwarf elliptical galaxies are the most common galaxy type in nearby galaxy clusters, yet they remain relatively poorly studied objects and many of their basic properties have yet to be quantified. In this contribution we present the preliminary results of a study of 4 Virgo and 1 field galaxy obtained with the SAURON integral field unit on the William Herschel Telescope (La Palma). While traditional long-slit observations are likely to miss more complicated kinematic features, with SAURON we are able to study both kinematics and stellar populations in two dimensions, obtaining a much more detailed view of the mass distribution and star formation histories.
\end{abstract}
\maketitle

\section{Introduction}
Dwarf elliptical galaxies (dEs) are the most common galaxy type in the nearby universe, accounting, for about 75\% of all objects in the Virgo cluster (e.g. Trentham \& Tully 2002). As potential building blocks of massive galaxies (in the current hierarchical galaxy formation paradigm), dEs may provide important clues on the main processes involved in galaxy assembly and evolution. A detailed analysis of structural properties, internal kinematics and stellar populations is, therefore, essential to understanding the evolutionary properties of this class of objects.

\section{Observations}
\label{observations}
We observed four Virgo and one field galaxy during 3 nights in January 2010 using the SAURON instrument mounted on the 4.2m William Herschel Telescope at the Observatorio del Roque de los Muchachos in La Palma, Spain. The average exposure time was 5h per galaxy. Additionaly, we used the WHT auxiliary camera ACAM to obtain deep images (300s) for each object.

\section{Stellar kinematics}
We used the penalized pixel fitting (pPXF) method of Capellari and Emselem (2004) to derive stellar absorption
line kinematics for each galaxy by directly fitting the spectra in the pixel space. We used absorption
spectra taken from the single-burst stellar population models of Vazdekis (1999).
    
\begin{figure*}
\centering
\resizebox{0.7\columnwidth}{!}{%
  \includegraphics{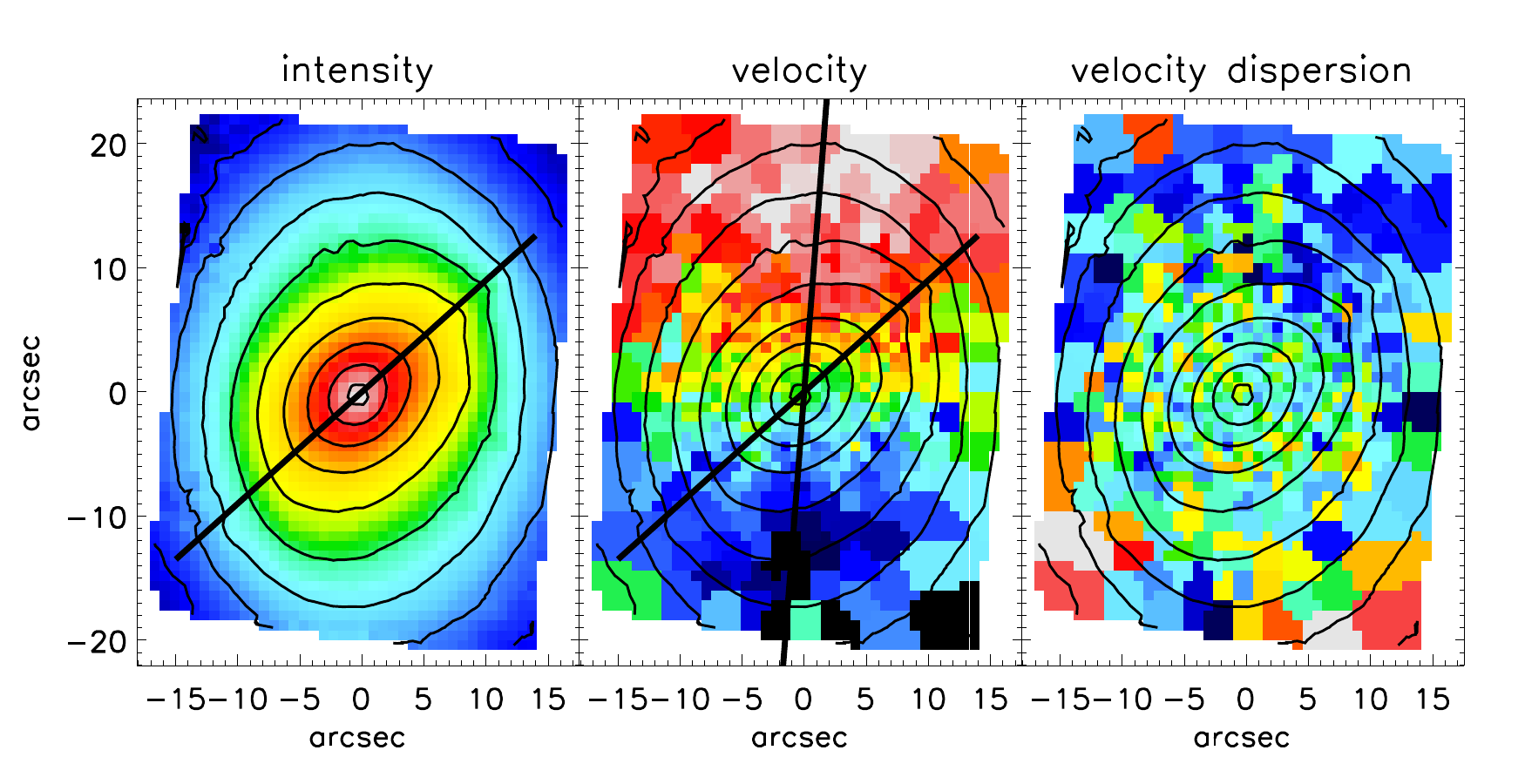}}
\caption{\textit{Maps of reconstructed intensity, $V$ and $\sigma$ for VCC 523. The galaxy is known to host a bar (Lisker et al. 2006), which is evidenced here by the misalignment between the photometric and kinematic major axes, overplotted on the intensity and the $V$ maps.}}
\label{map}       
\end{figure*}

We find one rotating galaxy with misaligned photometric and kinematic major axes, indicating the presence of a bar (see Fig.~\ref{map}). Our field galaxy, NGC 3073, also shows signs of rotation. Two out of three of our objects that do not appear to rotate are significantly flattened (VCC 1087 and VCC 1261), we speculate that they might be triaxial systems.

\section{Star formation histories}
We used the ULySS full spectrum fitting package of Koleva et al. (2009) together with Pegase.HR/ELodie.3.1 models (Prugniel et al. 2007) to estimate ages and metallicities ($Z$) of our objects. We fitted 1-SSP models to obtain equivalent ages and $Z$ and found significant gradients in at least one of the objects (VCC 1261).



\begin{thebibliography}{99}

\bibitem[2001]{bacon} Bacon, R. \etal\ 2001, MNRAS, 326, 23.
\bibitem[2004]{cappellari} Cappellari, M. \etal\ 2004, PASP, 116, 138.
\bibitem[2009]{koleva} Koleva M. \etal\ 2009, A\&A, 501, 1269.
\bibitem[2006]{lisker} Lisker, T. \etal\ 2006, AJ, 132, 497. 
\bibitem{prugniel} Prugniel, P. \etal\ VizieR On-line Data Catalog: III/251.
\bibitem{trentham} Trentham, N. \& Tully, R.~B. 2002, MNRAS, 335, 712.
\bibitem[1999]{vazdekis1} Vazdekis, A. \etal\ 1999, AJ, 513, 224.

\end{thebibliography}
\end{document}